\documentclass[epj,nopacs]{svjour}
\usepackage{graphics}
\usepackage{amsmath}
\usepackage{pdfpages}
\usepackage{cite}
\usepackage[hyphens]{url}
\usepackage[colorlinks=true,linkcolor=blue,citecolor=blue]{hyperref}%

\begin{document}
\title{Quark Models and Radial Oscillations: Decoding the HESS J1731-347 Compact Object's Equation of State}
\author{Ishfaq A. Rather\inst{1}  \and Grigoris Panotopoulos\inst{2} \and Il{\'i}dio Lopes \inst{1}
\authorrunning{Ishfaq et al.}
\thanks{\emph{Email:} ishfaq.rather@tecnico.ulisboa.pt (corresponding author)}%
}                     
\offprints{}          
\institute{Centro de Astrof{\'i}sica e Gravita{\c c}{\~a}o-CENTRA, Instituto Superior T{\'e}cnico,
Universidade de Lisboa, 1049-001 Lisboa, Portugal \and Departamento de Ciencias F{\'i}sicas, Universidad de la Frontera, Casilla 54-D, 4811186 Temuco, Chile} 
\date{Received: date / Revised version: date}
%
\abstract{
We investigate the peculiar nature of strange stars through an analysis of different quark models, i.e. vBag model and CFL model equation of states at different parameter sets, and focus on understanding the equation of state governing the intriguing central compact object (CCO) within the supernova remnant HESS J1731-347, with a mass and radius of $M = 0.77^{+0.20}_{-0.17} M_{\odot}$ and $R = 10.4^{+0.86}_{-0.78}$ km, respectively. Additionally, we compare the radial oscillations of two models to determine the frequency of the HESS J1731-347 compact object at its maximum mass. The frequencies of radial oscillations are computed for each of the four EoSs considered.  In total, the 10 lowest radial frequencies for each of those  EoSs have been computed. By delving into these aspects, we aim at deepening our understanding of strange stars and their connection to the observed HESS J1731-347 mass-radius relationship.
\PACS{
      {PACS-key}{discribing text of that key}   \and
      {PACS-key}{discribing text of that key}
     } 
} 
\authorrunning{Ishfaq \textit{et al.}}
\titlerunning{HESS J1731-34 Equation of State}
\maketitle
\section{Introduction}\label{sec1}

Ever since Witten proposed that quark matter may be the true ground state of hadronic matter \cite{Witten:1984rs}, followed by Bodmer's important precursor \cite{Bodmer:1971we}, a substantial amount of work has been done. The study of less standard stars, known as Strange Stars (SSs), has been inspired by the hypothesis that strange matter, also known as quark matter, composed of $u$, $d$, and $s$ quarks, is characterized by energy per baryon lower than that of nuclear matter and $u$, $d$ quark matter, along with the expectation of deconfined quark matter at high densities \cite{Ivanenko:1965dg, Itoh:1970uw,Collins:1974ky,Weber:2004kj,Alcock:1986hz,Haensel:1986qb}. Due to their nature, those objects can have arbitrarily small radii and masses \cite{schaffner2020compact}.

Those extremely dense, compact stars offer a unique potential to investigate the properties of matter under extreme conditions. Due to their unusual inner composition, structure, and physical properties compared to other compact objects, such as white dwarf (WD) or neutron stars (NSs) \cite{1983bhwd.book.....S}, such stars (i.e. SSs) have emerged as a topic of considerable interest over the last few decades.

The physics of very high-density matter, such as SSs, is still not fully understood. An Equation of State (EoS), or the relationship between pressure $P$ and energy density $\mathcal{E}$, must be specified in order to build a compact star's model. Properties for a compact object, such as the mass-radius relationship, tidal deformability, the rate of cooling, etc., can be provided by a clear EoS. The most basic MIT Bag model EoS framework is employed in the majority of SS studies. It is considered that a constant pressure, known as the Bag constant, on the surface of any region containing quarks essentially causes quark confinement in the MIT Bag model EoS, which is the simplest EoS and corresponds to a relativistic gas of de-confined quarks with energy density \cite{Haensel:1986qb,PhysRevD.9.3471}.

The observed massive compact objects, however, cannot be produced by the EoS of matter evaluated from this model because it is too soft. By incorporating a perturbative correction with a non-zero strong coupling constant, $\alpha_c$, the original bag model is modified. Implementing vector interaction between quarks by coupling to a vector field is another approach to incorporating quark interactions. If the vector interaction is repulsive, the EoS will become stiff, increasing its possible maximum mass. This model has been discussed in numerous works and is commonly known as the vector
interaction enhanced bag (vBag) model \cite{Kl_hn_2015, Rather_2021, PhysRevC.103.055814, Rather:2022bmm,Lopes:2020btp,vbageos}.

The aforementioned matter could also be a color superconductor, a degenerate Fermi gas in which the quarks close to the Fermi surface form Cooper pairs, breaking the color gauge symmetry \cite{alford2001color,RevModPhys.80.1455}. The color-flavor-locked (CFL) phase, a superfluid that also violates chiral symmetry, is the most favorable state for strange quark matter at asymptotically high densities.  Since all quarks in this state have the same Fermi momentum and cannot contain electrons, quarks form Cooper pairs of various colors and flavors in this state \cite{alford1999color,alford2001color}. Several properties of Quark matter, including its transport properties, are significantly impacted by color-flavor locking. For typical values of the color superconducting gap ($\Delta$ $\sim$ 0-150 MeV) and the baryon chemical potential ($\mu$ $\sim$ 300-400 MeV), it introduces corrections of order ($\Delta$/$\mu$)$^2$, which is roughly a few percent, into the EoS. However, in the low-pressure regime where quark matter's absolute stability is affected, the effect is proportionally very large. 
Thus, for a wide range of parameters of the MIT bag model EoS, self-bound stars made up of quark matter from the stellar core up to the stellar surface (SSs) may exist \cite{PhysRevD.66.074017}. Studies of the structure of these objects reveal that the mass-radius relationship of strange stars is significantly impacted by color superconductivity, allowing for very large maximum masses \cite{lugones2003high,horvath2004self}.

Apart from the vBag and the CFL model, other phenomenological quark models such as density dependent quark mass \cite{Fowler1981,CHAKRABARTY1989112,Chu_2014}, quasi particle \cite{SCHERTLER1997659,SCHERTLER1998451} ,the Nambu Jona-Lasinio (NJL) \cite{PhysRev.122.345,PhysRev.124.246} model have been widely. While the NJL model lacks confinement, it presents the chiral symmetry. Its extension, Polyakov–Nambu–Jona–Lasinio (PNJL) model \cite{FUKUSHIMA2004277,PhysRevD.77.114028} includes effects of confinement / deconfinement phase transition. The quarkyonic matter has been used in recent studies \cite{PhysRevD.102.023021,PhysRevLett.122.122701} which serves as an alternate approach that explicitly implements early theories about quark matter through the fact that both quarks and nucleons exist as quasiparticles in a crossover transition. However the recent observations and estimates of the mass-radius region for the
supermassive compact stars set very strict constraints on the EoS of strange quark matter (SQM) and may rule out most of the conventional phenomenological models of quark matter.

An indirect approach to studying such highly compact stars is asteroseismology \cite{2013pss4.book..207H}. One can identify a star's stability condition, mass, radius, composition, etc. by observing the radial and non-radial oscillations of the star. Even though the radial mode is the simplest oscillation mode and involves a regular change in the oscillating body's size and shape, it is a great tool for acquiring information about the star.

 In a pioneering work, Chandrasekhar \cite{PhysRevLett.12.114, 1964ApJ...140..417C} investigated the radial oscillations of stellar models. Significantly, the characteristics of the radial oscillation can provide information about the stability and EoS of compact stars. The non-radial oscillations of relativistic stars were studied by  Thorne and Campolattaro \cite{1967ApJ...149..591T}. They are known as quasinormal modes (QNMs) because the GWs dampen the oscillation modes. Typical nonrotating relativistic fluid stars are categorized as QNMs along the polar and axial axes. Among the polar modes are the fundamental ($f$), pressure ($p$), and gravity ($g$) modes. The axial modes only contain the space-time ($w$) modes. Radial oscillations are unable to generate GWs on their own, making it challenging to detect them \cite{PhysRevD.73.084010,PhysRevD.75.084038}. They are connected to non-radial oscillations that amplify GWs and enhance the likelihood of detecting them \cite{PhysRevD.73.084010,PhysRevD.75.084038}. Chirenti {\textit{et al.}} \cite{Chirenti_2019} observed that a hyper-massive NS formed in the post-merger event of a BNS along with the emission of a short gamma-ray burst (SGRB), which may be influenced by radial oscillations. It was feasible to see the hyper-massive NS's high-frequency oscillations, which ranged from $(1-4) ~ $kHz.

The fundamental and first two excited modes of the radial oscillations of zero-temperature degenerate stars (WD and NS) were calculated by Chanmugam \cite{1977ApJ...217..799C}. Haensel {\textit{et al.}} \cite{1989AA...217..137H} investigated the pulsation properties of NSs undergoing a phase transition to quark matter using the polytropic model EoS. Datta \cite{DATTA1992313} calculated the eigenfrequencies of radial pulsations for SSs using the general relativistic pulsation equation. The oscillation time periods were also computed. Vaeth and Chanmugam \cite{1992AA...260..250V} reported on the calculation of the radial oscillations of SSs and NSs for the two lowest-order oscillation modes.

The $f$-mode oscillations of NSs are very significant, as they are most likely to be detectable with a third-generation detector, such as the Einstein Telescope and the Cosmic Explorer \cite{2019BAAS...51c.251S, Punturo_2010, 2021arXiv211106990K}, or even in the best case by the current generation LIGO/Virgo/KAGRA detectors \cite{PhysRevLett.122.061104, 10.1093/ptep/ptac073}, as it depends on the EoS of compact stars. It is also anticipated to be excited in many astrophysical scenarios and hence result in efficient emission of gravitational waves.

According to a recent analysis of the supernova remnant HESS J1731-347, the central compact object (CCO) mass and radius are $M = 0.77^{+0.20}_{-0.17} M_{\odot}$ and $R = 10.4^{+0.86}_{-0.78}$ km, respectively \cite{Doroshenko2022}.
This estimate of maximum mass, $0.77 ~ M_{\odot}$, is intriguing. It is possible that the compact object in HESS J1731-347 is an exotic object rather than a neutron star, given that \cite{suwa2018minimum}  previous analysis revealed that the minimum possible mass of a neutron star is 1.17$M_{\odot}$. Several studies have already shown that the compact object in HESS J1731-347 could be a neutron star or a strange star \cite{li2023baryonic,sagun2023nature,oikonomou2023colourflavour,huang2023hadronic,Das:2023qej,Chu2023,PhysRevC.108.025808}.

In the present work, we employ two different quark models, the vBag and CFL, to study SSs. We vary the vector coupling parameter $K_v$ in the vBag model and the color superconducting gap parameter $\Delta$ in the CFL model to produce four different EoSs in total. Our main goal
is to present the capability of vBag and CFL quark stars to model
the HESS J1731-347 compact object, as well as objects
with masses equal or greater than 2 $M_{\odot}$ limit. We also investigate various radial oscillations of SSs with different EoSs at the mass corresponding to the maximum mass of HESS J1731-347, $M$ = 0.77 $M_{\odot}$, to obtain the frequencies that could be emitted by this CCO, if it's a SS. Several studies on the investigation of various radial oscillations of NSs with different exotic phases have already been carried out \cite{10.1093/mnras/stac2622, kokkostas, https://doi.org/10.48550/arxiv.2205.02076, PhysRevD.101.063025, PhysRevD.98.083001, https://doi.org/10.48550/arxiv.2211.12808, PhysRevD.107.123022}. Moreover, as already mentioned before, several studies have shown the nature of the compact object in HESS J1731-347 to be either an NS or a SS, but none among them have calculated the radial oscillations.

 Our work is organized as follows: in Sec. (\ref{sec:eos1}) and (\ref{sec:eos2}), the EoS for the vBag and the CFL model will be discussed. The tidal Love numbers and deformability are discussed in Sec. (\ref{tidal}), while the Sturm-Liouville eigenvalue problem for the inner structure and radial oscillations of SSs are introduced in Sec. (\ref{radial}), respectively. In Sec. (\ref{results}) the trace anomaly and the Mass-Radius profile for different EoSs are discussed in Sec. (\ref{mr}). Sec. (\ref{profile}) describes the numerical results for radial profiles obtained for SSs. The summary and concluding remarks are finally given in Sec. (\ref{summary}).

\section{Theoretical framework and formalism}

\subsection{Quark models and equations-of-state}

\subsubsection{vBag model}\label{sec:eos1}

Quark matter in NSs has been extensively described using the MIT bag model. Quarks are interpreted as being free inside a bag in the original description, and thermodynamic properties are simply derived from a free Fermi gas model \cite{PhysRevD.30.2379,PhysRevD.9.3471,PhysRevD.17.1109}. A more useful model for investigating astrophysical processes is the vBag model \cite{Kl_hn_2015}, a modified version of the bag model. It is preferred over the simple bag model because it accounts for repulsive vector interactions as well as dynamic chiral symmetry breaking (D$\chi$SB). The significance of the repulsive vector interaction stems from the fact that it enables the pure strange stars to reach and satisfy the $2$ $M_{\odot}$ maximum mass limit \cite{vbageos,Rather_2021,Lopes:2020btp, Antoniadis1233232}.

The Lagrangian density for the vBag model along with a free Fermi gas of leptons is written as
\begin{equation} \label{eq1}
\begin{split}
\mathcal{L} &=\sum_f  [ \psi_f (i\gamma_{\mu} \partial_{\mu} -m_f -B_{bag}) \psi_f] \Theta_H  \nonumber \\
& -G_V \sum_f (\bar{\psi_f} \gamma_{\mu} \psi_f)^2
+ \sum_l  \psi_l \gamma_{\mu} (i \partial_{\mu} -m_l )\psi_l \ ,
\end{split}
\end{equation}
where $f$ = $u$, $d$, $s$ denotes the quarks and $l$ = $e^-$, $\mu^-$ represents leptons. $B_{bag}$ denotes the bag constant and $\Theta_H$ is the Heaviside step function which allows for the confinement/deconfinement of the bag \cite{PhysRevD.30.2379}. The vector interaction coupling constant $G_V$ is introduced via the coupling of the vector-isoscalar meson to the quarks.

The total energy density and pressure are
\begin{equation}
\mathcal{E}_Q = \sum_{l} \mathcal{E}_l+\sum_{f=u,d,s} \mathcal{E}_{\rm{vBag},f}-B_{dec}\ ,
\end{equation}
\begin{equation}
P_Q = \sum_{l} P_l+\sum_{f=u,d,s} P_{\rm{vBag},f}+B_{dec}\ ,
\end{equation}
where $B_{dec}$ represents the de-confined bag constant. The energy density and pressure of a single quark flavor are defined as
\begin{equation}\label{q1}
\mathcal{E}_{\rm{vBag},f} = \mathcal{E}_f(\mu_f^*)+\frac{1}{2}K_{\nu}n_f^2 (\mu_f^*)+B_{\chi,f}\ ,
\end{equation}   
\begin{equation}
P_{\rm{vBag},f}= P_f(\mu_f^*)+\frac{1}{2}K_{\nu}n_f^2 (\mu_f^*)-B_{\chi,f}\ .
\end{equation}
 The coupling constant parameter $K_{\nu}$ = 2$G_V$ results from the vector interactions and controls the stiffness of matter \cite{Wei_2019}. The bag constant for a single quark flavor is denoted by $B_{\chi,f}$ and the effective bag constant $B_{\rm{eff}}$ is defined as 
\begin{equation}
B_{\rm{eff}}=\sum_{f=u,d,s}B_{\chi,f}-B_{dec}\ .
\end{equation}

The effective chemical potential $\mu_f^*$ of the system and the quark density are defined as
\begin{equation}
\mu_f^* =\mu_f -K_{\nu}n_f(\mu_F^*)\ ,
\end{equation}  
\begin{equation}
n_f (\mu_f) = n_f (\mu^*)\ .
\end{equation}

In the present study, we will use two values of the coupling constant parameter $K_{\nu}$ = 6 and 9 GeV$^{-2}$. The effective bag constant $B_{eff}$ will be kept constant at $B_{eff}$ = 55 MeV/fm$^{3}$. The parameters $K_{\nu}$ and $B_{eff}$ are chosen so as to satisfy the stability limit and to remain consistent with the 2 $M_{\odot}$ constraint on the maximum mass. Lower values of the coupling parameter and higher values of the effective bag constant don't satisfy this 2 $M_{\odot}$ limit on the maximum mass and hence the EoSs produced are ruled out. Increasing the coupling constant parameter while keeping the bag constant fixed ensures a very stiff EoS that satisfies all the mass and radius constraints from various observations. For simplicity, we will use $B$ instead of $B_{eff}$ throughout the paper. The EoS at $B$ = 55 $\mathrm{ MeV/fm^3}$, $K_{\nu}$ = 9 $\mathrm{GeV^{-2}}$ will be denoted by vBag1, while the EoS at $B$ = 55 $\mathrm{MeV/fm^3}$, $K_{\nu}$ = 6 $\mathrm{GeV^{-2}}$ will be denoted by vBag2. \par

The charge neutrality and chemical-equilibrium conditions for the quark matter are
\begin{equation}
\frac{2}{3}\rho_u -\frac{1}{2}(\rho_d+\rho_s)-\rho_e-\rho_\mu =0\ ,
\end{equation}
\begin{align}
&\mu_s=\mu_d=\mu_u+\mu_e \\
&\mu_\mu=\mu_e.
\end{align}

\subsubsection{CFL model}\label{sec:eos2}

The expression for energy density and pressure for the CFL quark matter to order $\Delta^2$ and $m_s^2$ can be written as \cite{PhysRevD.66.074017, PhysRevC.95.025808}

\begin{equation}
    P = \frac{3\mu^2}{4\pi^2}+\frac{9\alpha \mu^2}{2\pi^2} -B\ ,
\end{equation}
\begin{equation}
    \mathcal{E} = \frac{9\mu^2}{4\pi^2}+\frac{9\alpha \mu^2}{2\pi^2} +B\ ,
\end{equation}
where
\begin{equation}
    \alpha = -\frac{m_s^2}{6}+\frac{2\Delta^2}{3}\ .
\end{equation}

An analytic expression for $\mathcal{E} = \mathcal{E}(P)$ can be obtained from the above expressions as
\begin{equation}
    \mathcal{E} = 3P+4B-\frac{9\alpha \mu^2}{\pi^2}\ ,
\end{equation}
with
\begin{equation}
    \mu^2  = -3\alpha+\Bigg[\frac{4}{3}\pi^2 (B+P)+9\alpha^2 \Bigg] ^{1/2}\ .
\end{equation}

Similarly for $P = P(\mathcal{E})$, we obtain
\begin{equation}
    P = \frac{\mathcal{E}}{3}-\frac{4B}{3}+\frac{3\alpha \mu^2}{\pi^2}\ ,
\end{equation}
with
\begin{equation}
    \mu^2  = -\alpha+\Bigg[\frac{4}{9}\pi^2 (\mathcal{E}-B)+ \alpha^2 \Bigg] ^{1/2}\ .
\end{equation}

Since the values of $\Delta$, $B$, and $m_s$ are not accurately known, they are considered free parameters. For absolutely stable strange matter, the energy per baryon of CFL quark matter must be less than the neutron mass $m_n$ at zero pressure and temperature.
As shown in Ref. \cite{PhysRevC.95.025808} considering the stability window, several EoS with different values of $\Delta$, $B$, and $m_s$ have been studied. For the current study, we choose two different sets $(B, \Delta, m_s)$ = (60, 150, 150) and (60, 100, 0). These two produce a very stiff quark EoS with a maximum mass in the range of 2.3 - 2.6 $M_{\odot}$ thus satisfying the 2$M_{\odot}$ limit as well. 
The CFL EoS at $B$ = 60 $\mathrm{MeV/fm^3}$, $\Delta$ = 150 $\mathrm{MeV}$, $m_s$ = 150 $\mathrm{MeV}$ will be denoted by CFL1 whereas the EoS at $B$ = 60 $\mathrm{MeV/fm^3}$, $\Delta$ = 100 $\mathrm{MeV}$, $m_s$ = 0 $\mathrm{MeV}$ will be denoted by CFL2.

\subsection{Tidal Love numbers and deformablity}
\label{tidal}

The relativistic theory of tidal Love numbers for compact objects can be found in \cite{Hinderer:2007mb,Damour:2009vw,Postnikov:2010yn}. Here we briefly summarize the main formulas we shall be using throughout the present article.

The dimensionful deformability, $\lambda_{tidal}$, and the dimensionless one, $\Lambda$, are computed in terms of the tidal Love number, $k$, as follows
\begin{eqnarray}
\lambda_{tidal} & = & \frac{2 k R^5}{3}\ , \\
\Lambda & = & \frac{2 k}{3 C^5}\ ,
\end{eqnarray}
where $C=M/R$ is the factor of compactness. Next, in terms of $C$ the tidal Love number is computed to be
\begin{equation}
    \displaystyle k = \frac{8 C^5}{5} \:  \frac{K_o}{3 K_o \ln (1 - 2 C) + P_5(C)}\ ,
\end{equation}
where $K_o$ and $P_5(C)$ depend on $C, y_R$ and they are found to be
    \begin{equation}
        K_o = (1 - 2 C)^2 \; [2 C (y_R - 1) - y_R + 2]\ ,
    \end{equation}
    \begin{equation}
        y_R \equiv y (r = R)\ ,
    \end{equation}


with $P_5(C)$ being a fifth-order polynomial given by 
    \bigskip
    \begin{align}
         \displaystyle P_5(C) & = 2C \; [ 4C^4 (y_R + 1) + 2C^3 (3 y_R - 2) \nonumber \\
         & + 2C^2 (13 - 11 y_R) + 3C (5 y_R - 8) - 3 y_R + 6 ]\ ,
    \end{align}

    
while the function $y(r)$ satisfies a Riccati differential equation
    \bigskip
    \begin{align}
        \displaystyle r y'(r) + y(r)^2 + y(r) e^{\lambda (r)} [1 & + 4 \pi r^2 ( p(r) - \rho (r) ) ] \nonumber \\
        & + r^2 Q(r) = 0\ ,
    \end{align}

and where 
\begin{align}
        \displaystyle Q(r) & = 4 \pi e^{\lambda (r)} \left[ 5 \rho (r) + 9 p(r) + \frac{\rho (r) + p(r)}{c^2_s(r)} \right] \nonumber \\
        & - 6 \frac{e^{\lambda (r)}}{r^2} - [\nu' (r)]^2\ ,
    \end{align}

    \bigskip

is given in terms of known functions of the internal solution. The metric functions $\lambda$ and $\nu$ are defined as 
\begin{equation}
	e^{2\nu(r)} = e^{-2\lambda(r)} = 1-2m/r
\end{equation}
The interested reader may consult, e.g., \cite{Hinderer:2007mb,Damour:2009vw,Postnikov:2010yn} for more details about the tidal properties.

\subsection{Radial oscillations}
\label{radial}

The radial oscillation properties can be determined using the static equilibrium structure-based Einstein field equation \cite{1966ApJ...145..505B}. The metric is now time-dependent in a spherically symmetric system with only radial motion.
 The small perturbation of the equations for radial displacement $\Delta r$ with $\Delta P$ as the perturbation of the pressure, governing the dimensionless quantities $\xi$ = $\Delta r/r$ and $\eta$ = $\Delta P/P$ are defined as \cite{1977ApJ...217..799C, 1997AA...325..217G}
 \begin{equation}\label{ksi}
     \xi'(r) = -\frac{1}{r} \Biggl( 3\xi +\frac{\eta}{\gamma}\Biggr) -\frac{P'(r)}{P+\mathcal{E}} \xi(r)\ ,
 \end{equation}
 \begin{equation}\label{eta}
 \begin{split}
          \eta'(r) = \xi \Biggl[ \omega^{2} r (1+\mathcal{E}/P) e^{\lambda - \nu } -\frac{4P'(r)}{P} -8\pi (P+\mathcal{E}) re^{\lambda} \\
     +  \frac{r(P'(r))^{2}}{P(P+\mathcal{E})}\Biggr] + \eta \Biggl[ -\frac{\mathcal{E}P'(r)}{P(P+\mathcal{E})} -4\pi (P+\mathcal{E}) re^{\lambda}\Biggr]\ ,
      \end{split}
 \end{equation}
 where $\omega$ is the frequency oscillation mode and $\gamma$ is the adiabatic relativistic index defined as
 \begin{equation}
     \gamma = \Biggl( 1+\frac{\mathcal{E}}{P}\Biggr) v_s^{2}\ ,
 \end{equation}
 where $v_s^{2}$ is the speed of sound squared
  \begin{equation}\label{cs}
     v_s^{2} = \Biggl(\frac{dP}{d\mathcal{E}}\Biggr)c^{2}\ .
 \end{equation}

Eqs.~(\ref{ksi}) and (\ref{eta}) are two coupled differential equations supplemented with two additional boundary conditions, one at the center where $r$ = 0, and another at the surface where $r$ = $R$. At the center, the boundary condition requires that
\begin{equation}
    \eta = -3\gamma \xi\ ,
\end{equation}
must be satisfied. Also, Eq.~\eqref{eta} must be finite at the surface, and hence
\begin{equation}
    \eta = \xi \Biggl[ -4 +(1-2M/R)^{-1} \Biggl( -\frac{M}{R} -\frac{\omega^{2} R^{3}}{M}\Biggr)\Biggr]\ ,
 \end{equation}
 must be satisfied where $M$ and $R$ correspond to the mass and radius of the star, respectively. The frequencies are computed by
 \begin{equation}
\nu = \frac{\bar{\omega}}{2\pi}~~(kHz)\ ,
\end{equation}
where $\bar{\omega}$ = $\omega t_0$ is the dimensionless quantity computed at $t_0$ = 1 ms.  

 These equations represent the Sturm-Liouville eigenvalue equations for $\omega$. The solutions provide the discrete eigenvalues $\omega_n ^{2}$ and can be ordered as 
 \begin{equation*}
\omega_0 ^{2} < \omega_1 ^{2} <... <\omega_n ^{2}\ , 
 \end{equation*}
 where $n$ is the number of nodes for a given star. The star will be stable for a real value of $\omega$ and unstable for an imaginary frequency. Additionally, because the eigenvalues are arranged in the manner previously mentioned, it is crucial to understand the fundamental $f$-mode frequency ($n$ = 0) in order to assess the star's stability.

\section{Numerical results and discussion}
\label{results}

\subsection{Trace Anomaly and MR Profile}
\label{mr}

The speed of sound in the case of neutron stars does not violate the conformal bound when $v_s^2$ $>$ 1/3, according to recent work \cite{PhysRevLett.129.252702, PhysRevC.107.025802}, which suggests the trace anomaly as a measure of conformality, but it exhibits a steep approach to the conformal limit. In particular, the trace anomaly is used exclusively to express the speed of sound $v_s$, and it is suggested that this makes it a more complete quantity than $v_s$. The definition of the normalized trace anomaly is
\begin{equation}
    D = \frac{1}{3}- \frac{P}{\mathcal{E}}\ .
\end{equation}
The original representation letter $\Delta$ is replaced by $D$ to avoid any confusion with the superconducting gap parameter.
Since the thermodynamic causality requires that $P$ $>$ 0 and $P$ $\le$ $\mathcal{E}$, the trace anomaly $D$ must satisfy the constraints -2/3 $\le$ $D$ $\le$ 1/3. 

The speed of sound can be written as
\begin{equation}\label{cs}
    v_s^2 = \frac{1}{3} -D-\frac{dD}{d\eta}\ ,
\end{equation}
where $\frac{1}{3} -D$ represents the non-derivative term and $-\frac{dD}{d\eta}$ represent the derivative term.  Here, $D_{log}$ = $ln(\mathcal{E}/\mathcal{E}_0)$ and $\mathcal{E}_0$ is the energy density at saturation density. Thus the Eq. (\ref{cs}) can be written as
\begin{equation}
   v_s^2 = v_{s, deriv}^2 + v_{s, nonderiv}^2\ .
\end{equation}

\begin{figure}
\centering
	\includegraphics[scale=0.32]{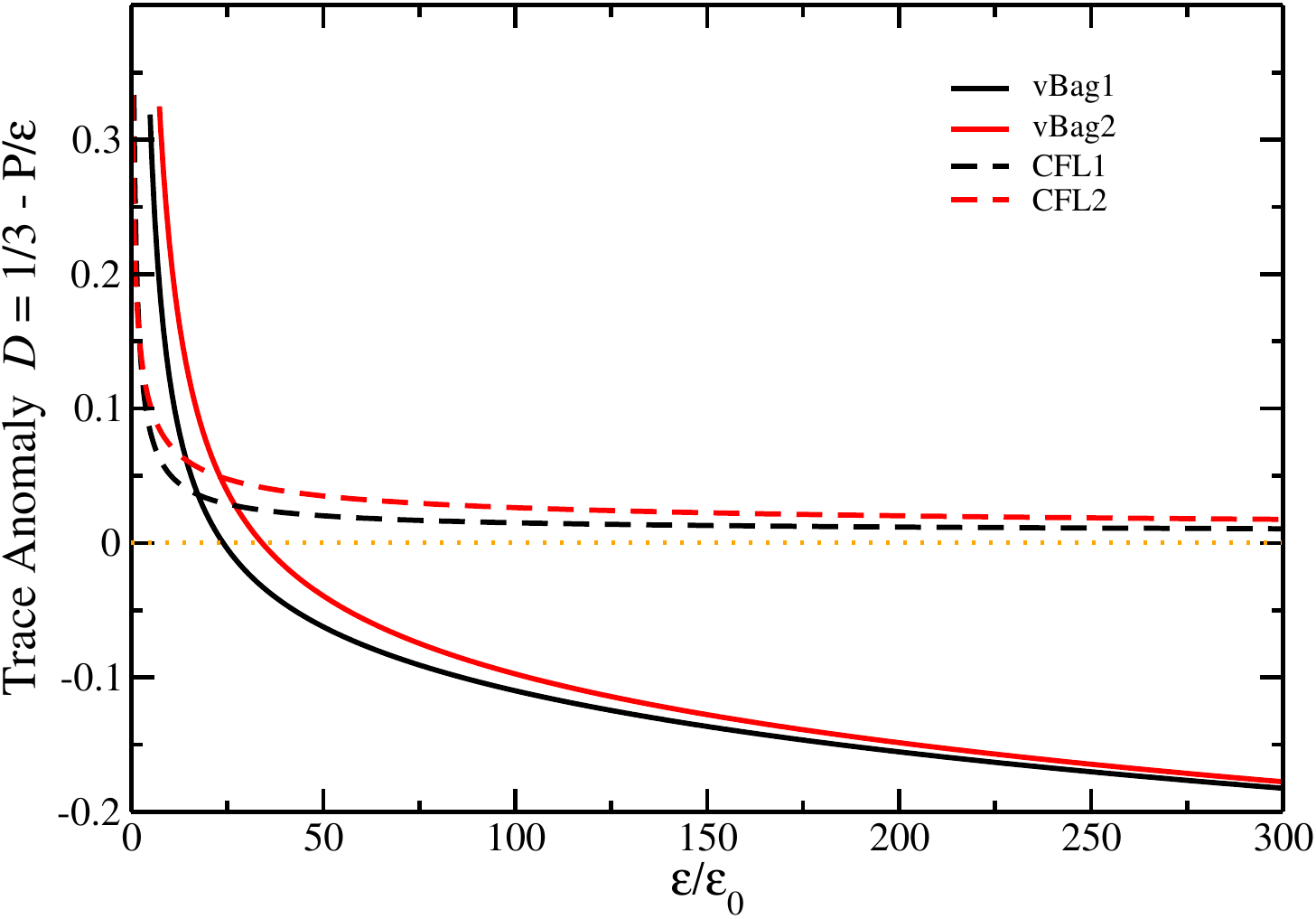}
	\caption{ Trace anomaly as a function of energy density (normalized to the value of the saturation density $\mathcal{E}_0$ = 159.4 MeV/fm$^3$) for $B$ = 55 $\mathrm{ MeV/fm^3}$, $K_{\nu}$ = 9 $\mathrm{GeV^{-2}}$ (vBag1) and $B$= 55 $\mathrm{MeV/fm^3}$, $K_{\nu}$ = 6 $\mathrm{GeV^{-2}}$ (vBag2). The results for the CFL EoS at $B$ = 60 $\mathrm{MeV/fm^3}$, $\Delta$ = 150 $\mathrm{MeV}$, $m_s$ = 150 $\mathrm{MeV}$ (CFL1) and CFL EoS at $B$ = 60 $\mathrm{MeV/fm^3}$, $\Delta$ = 100 $\mathrm{MeV}$, $m_s$ = 0 $\mathrm{MeV}$ (CFL2) are also shown.}
	\label{fig1} 
\end{figure}

Fig.~\ref{fig1} displays the trace anomaly $D$ as a function of normalized energy density $\mathcal{E}/\mathcal{E}_0$ for vBag and CFL model EoSs at different parameter sets. For the vBag model, the trace anomaly keeps decreasing with the increasing normalized energy density and drops below 0 at $\mathcal{E}/\mathcal{E}_0$ $\approx$ 23 and 34 for vBag1 and vBag2 EoS, respectively. For the CFL model, both CFL1 and CFL2 EoS decrease at lower values of the normalized energy density and then remain almost constant at higher values. The trace anomaly for both CFL EoSs doesn't become negative. At sufficiently high densities, as was initially demonstrated in Ref. \cite{Zeldovich:1961sbr}, the possibility of the negative trace anomaly is statistically not excluded. 
Even if the speed of sound develops a peak, the trace anomaly may still be monotonic \cite{PhysRevLett.129.252702}.

\begin{figure}
\centering
  	\includegraphics[scale=0.32]{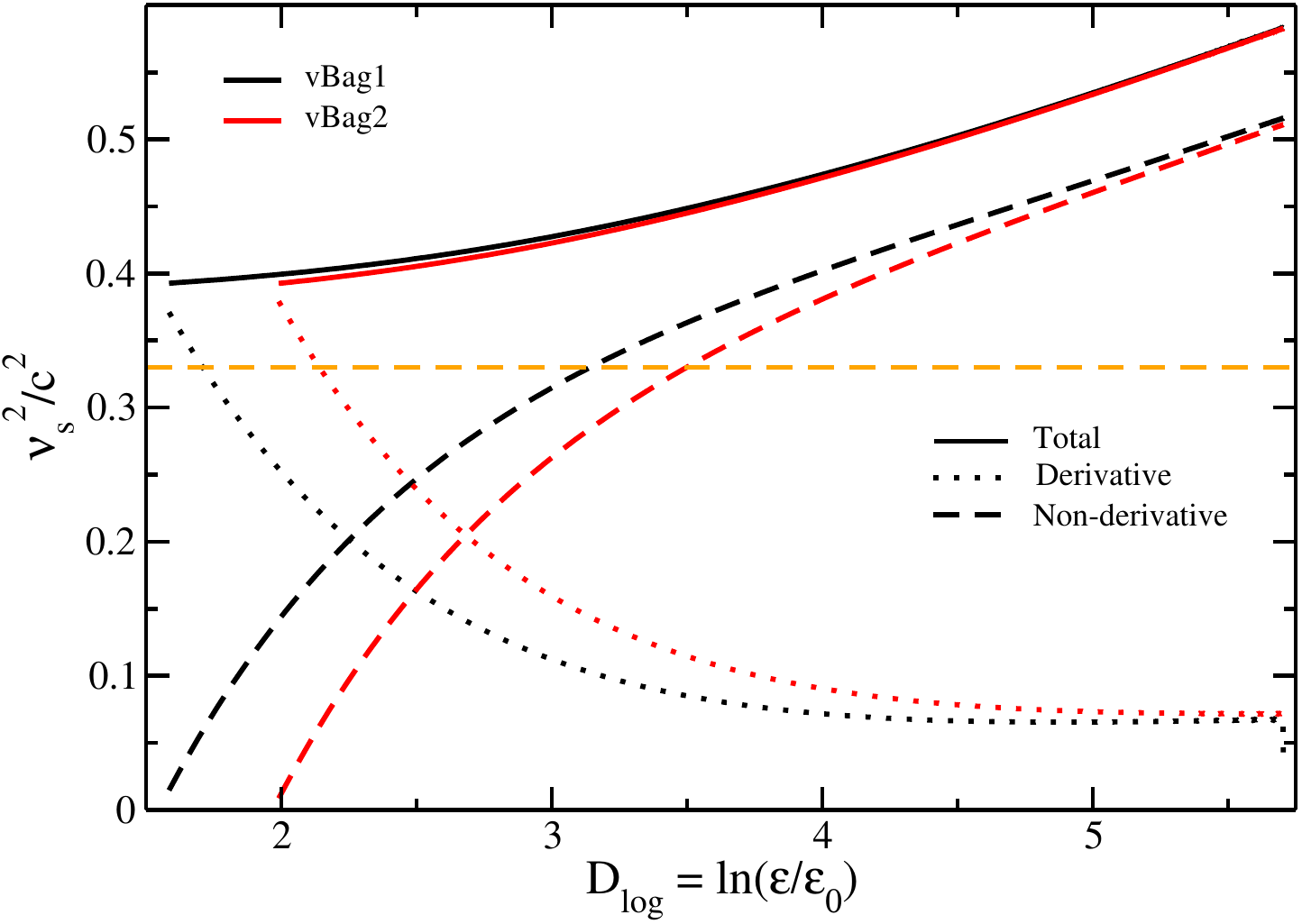}\\
     	\includegraphics[scale=0.32]{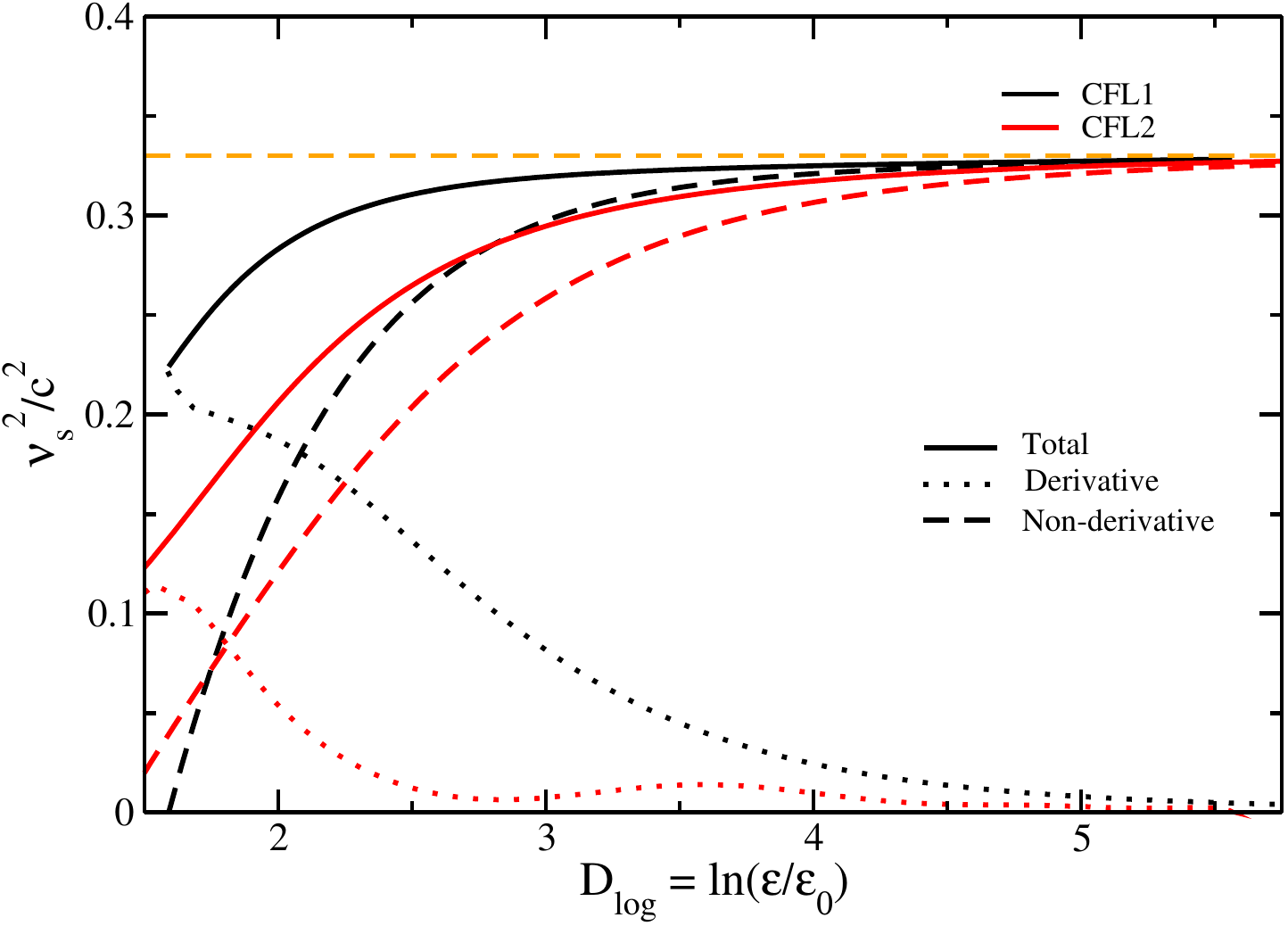}
  	\caption{ The speed of sound (solid line) along with the respective derivative (dotted line) and non-derivative terms (dashed line), as functions of the normalized logarithmic energy density $D_{log}$ = $ln(\mathcal{E}/\mathcal{E}_0)$ for vBag EoSs (upper figure) and for CFL EoSs (lower figure). The orange dashed line represents the conformal limit $c_s^{2}$ = 1/3.}
  	\label{fig2} 
  \end{figure}
  
An essential parameter that provides information regarding shear viscosity, tidal deformability, and gravitational wave signatures is the speed of sound \cite{PhysRevC.102.055801, Lopes_2021}.  It can also be interpreted as a measure of the stiffness of the EoS, with a higher speed resulting in a higher pressure at a given energy density and allowing a larger star mass for a given radius. 
 It is ensured by thermodynamic stability that $v_s^2$ $>$ 0 and by causality that $v_s^2$ $\le$ 1. According to perturbative QCD investigation, the upper limit for extremely high densities is $v_s^2$ = 1/3 \cite{PhysRevLett.114.031103}. According to several studies  \cite{PhysRevLett.114.031103, PhysRevC.95.045801, Tews_2018}, the 2 $M_{\odot}$ requirements demand a speed of sound squared that is greater than the conformal limit ($v_s^2$ = 1/3), indicating that the matter inside of NS is a highly interacting system.

Fig.~\ref{fig2} displays the speed of sound along with the respective derivative and non-derivative terms, as a function of the normalized logarithmic energy density $D_{log}$ =  $ln(\mathcal{E}/\mathcal{E}_0)$ for vBag and CFL model EoSs. For the vBag model EoSs, the speed of sound shows an increasing behavior and is higher than the conformal limit even at low values of the $D_{log}$. $v_{s, nonderiv}^2$ increases with the increasing $D_{log}$ because the trace anomaly $D$ is a monotonic function \cite{PhysRevLett.129.252702}. The $v_{s, deriv}^2$ shows the opposite behavior to $v_{s, nonderiv}^2$. It keeps decreasing up to $D_{log}$ $\approx$ 4.5 and remains almost constant thereafter. For CFL model EoSs, the speed of sound increases with the density but does not violate the conformal limit even at higher densities. 

 \begin{figure}
 \centering
 	\includegraphics[scale=0.32]{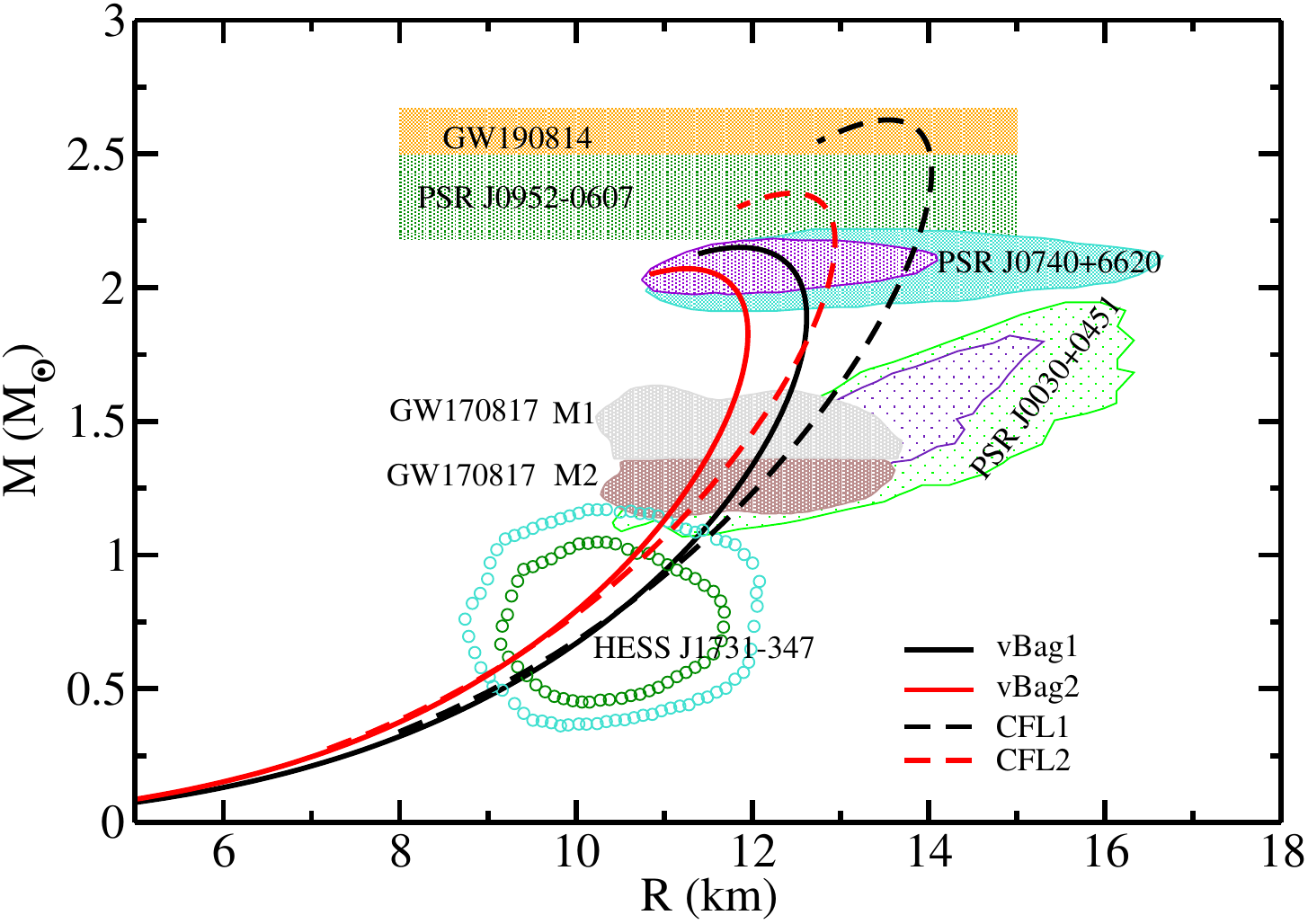}
 	\caption{ Mass-Radius profile for vBag EoS at $B$ = 55 $\mathrm{ MeV/fm^3}$, $K_{\nu}$ = 9 $\mathrm{GeV^{-2}}$ (vBag1) and $B$= 55 $\mathrm{MeV/fm^3}$, $K_{\nu}$ = 6 $\mathrm{GeV^{-2}}$ (vBag2). The results for the CFL EoS at $B$ = 60 $\mathrm{MeV/fm^3}$, $\Delta$ = 150 $\mathrm{MeV}$, $m_s$ = 150 $\mathrm{MeV}$ (CFL1) and CFL EoS at $B$ = 60 $\mathrm{MeV/fm^3}$, $\Delta$ = 100 $\mathrm{MeV}$, $m_s$ = 0 $\mathrm{MeV}$ (CFL2) are also displayed. The 68\% (violet) and 95\% (turquoise) credible regions for mass and radius are inferred from the analysis of PSR J0740+6620 \cite{miller2021,2021ApJ...918L..27R}. For PSR J0030+0451, the indigo dotted region is for 68\% credibility while the green dotted region is for 95\% credibility \cite{Miller_2019a}.   The grey upper (brown lower) shaded region corresponds to the higher (smaller) component of the GW170817 event \cite{Abbott_2020a}.  The circled contours represent the central compact objects within HESS J1731-347 \cite{Doroshenko2022}. The orange and green bands represent the maximum mass constraint from GW190814 data and PSR J0952-0607, respectively \cite{Abbott:2020khf,Romani_2022}.
  }
 	\label{fig3} 
 \end{figure}

Solving the Tolman–Oppenheimer–Volkoff (TOV) equations for the EoSs obtained, Fig.~\ref{fig3} displays the mass-radius profile for vBag and CFL EoS at different parameter sets. For vBag1 EoS with $B$ = 55 $\mathrm{ MeV/fm^3}$ and $K_{\nu}$ = 9 $\mathrm{GeV^{-2}}$, we obtain a maximum mass of 2.15 $M_{\odot}$ with a radius of 11.76 km. Similarly for vBag2 EoS with $B$ = 55 $\mathrm{ MeV/fm^3}$ and $K_{\nu}$ = 6 $\mathrm{GeV^{-2}}$, the maximum mass is 2.06 $M_{\odot}$ with a corresponding radius of 11.23 km. Both EoSs satisfy the PSR J0740+6620 upper mass limit.  The radius at 1.4 $M_{\odot}$ is 12.15 and 11.55 km, respectively, satisfying the constraints from
GW170817 and PSR J0030+0451.
For CFL1 EoS with $B$ = 60 $\mathrm{MeV/fm^3}$, $\Delta$ = 150 $\mathrm{MeV}$, $m_s$ = 150 $\mathrm{MeV}$, the maximum mass is 2.62 $M_{\odot}$ with radius of 13.50 km satisfying the GW190814 maximum mass constraint  \cite{Abbott:2020khf}. 
CFL2 EoS with $B$ = 60 $\mathrm{MeV/fm^3}$, $\Delta$ = 100 $\mathrm{MeV}$, $m_s$ = 0 $\mathrm{MeV}$ lies in the region of PSR J0952-0607 with maximum mass of 2.36 $M_{\odot}$. All these four EoSs satisfy the radius constraint of  $R$ = 10.4$_{-0.78}^{+0.86}$ km at mass $M$ = 0.77$_{-0.17}^{+0.20}$ $M_{\odot}$ from observables of HESS J1731-347 \cite{Doroshenko2022}.

\begin{center}
\begin{table}
		\caption{ Properties of SSs for four different EoSs, vBag1, vBag2, CFL1, and CFL2. $M_{max}$ represents the maximum mass with the corresponding radius $R_{max}$, $R_{1.4 M_{\odot}}$  and $R_{0.77 M_{\odot}}$ represent the radius at 1.4 and at 0.77 $M_{\odot}$, respectively. $\Lambda_{1.4 M_{\odot}}$  represents the dimensionless tidal deformability at 1.4 $M_{\odot}$.
  \label{table:eos} }
\begin{tabular}{p{0.8cm}p{0.8cm}p{0.8cm}p{0.9cm}p{1.0cm}p{1.0cm}p{1.0cm} }
 \hline
 EoS  & $M_{max}$  & $R_{max}$ & $R_{1.4 M_{\odot}}$  & $R_{0.77 M_{\odot}}$ & $\Lambda_{1.4 M_{\odot}}$ \\
 &  $(M_{\odot})$ & (km) & (km) & (km) & \\
 \hline
vBag1 & 2.15 & 11.76 & 12.15 & 10.43 & 158 \\
vBag2 & 2.06 & 11.23 & 11.55 & 9.90 & 119 \\
CFL1 & 2.62 & 13.50 & 12.47 & 10.47 & 484 \\
CFL2 & 2.36 & 12.39 & 11.86 & 10.05 & 212 \\
 \hline
\end{tabular}
\end{table}
\end{center}

\begin{figure}
	\includegraphics[scale=0.32]{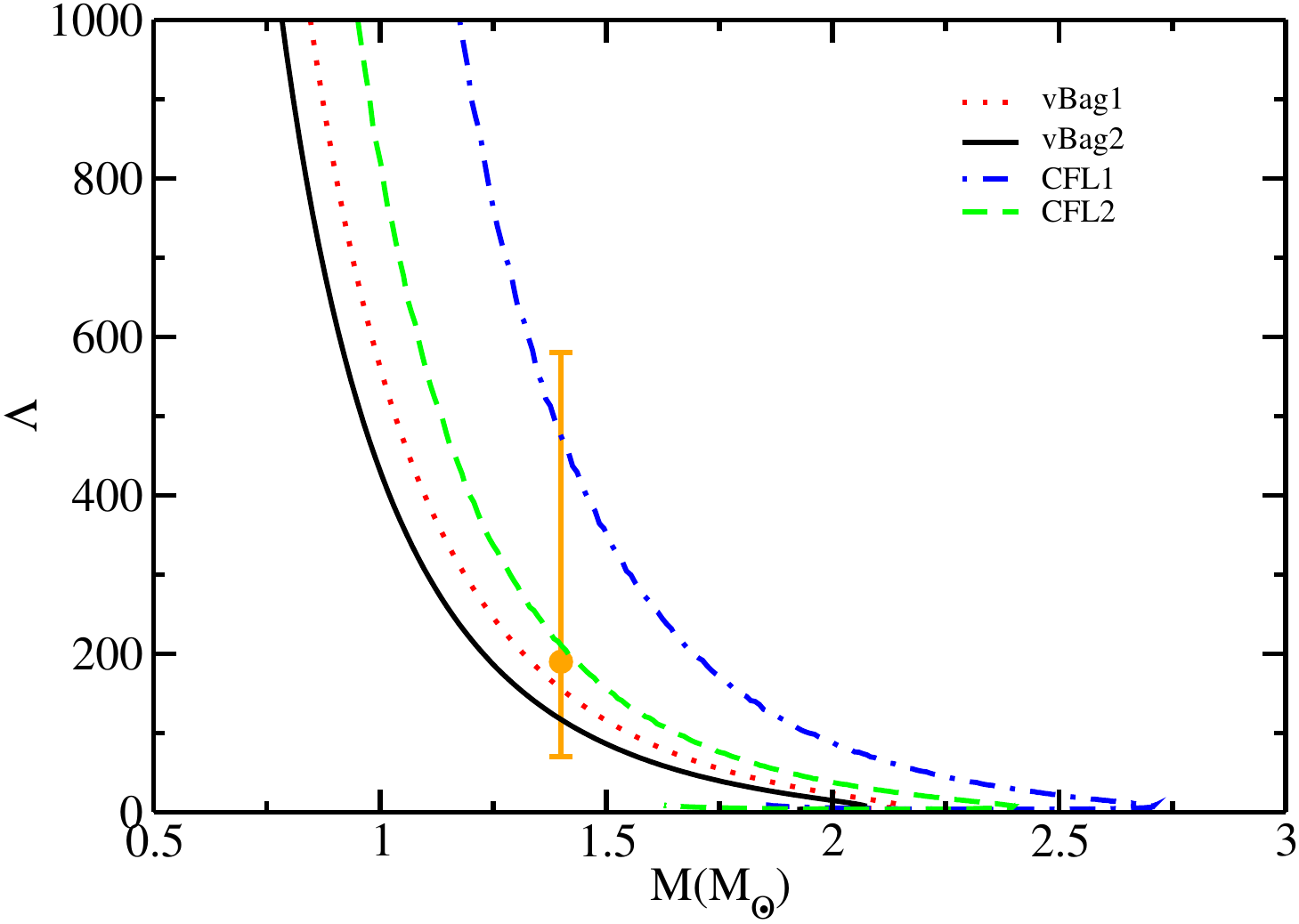}
	\caption{ (Color online) Dimensionless tidal deformability variation with the SS mass for vBag and CFL EoS at different parameter sets. The error bar represents the $\Lambda$ = 190$^{+390}_{-120}$ constraints concluded by the
 LVC measurement from GW170817 \cite{GW170817}. }
	\label{fig:tidal} 
\end{figure}

Fig.~\ref{fig:tidal} shows the dimensionless tidal deformability $\Lambda$ as a function of mass for different EoSs studied. For vBag1 and vBag2 EoS, the dimensionless tidal deformability at the canonical mass is around 115 and 163, respectively. Similarly for CFL1 and CFL2 EoS, the value of $\Lambda$ is around 250 and 480, respectively. All the values lie well within the constraint from GW170817, $\Lambda$ = 190$^{+390}_{-120}$ \cite{GW170817}. Table \ref{table:eos} displays the properties such as Maximum mass, corresponding radius, the radius at 1.4 and 0.77 $M_{\odot}$, and dimensionless tidal deformability, for all four EoSs.

\subsection{Radial Profiles}
\label{profile}

\begin{figure*}
\centering
    \includegraphics[width=0.47\linewidth]{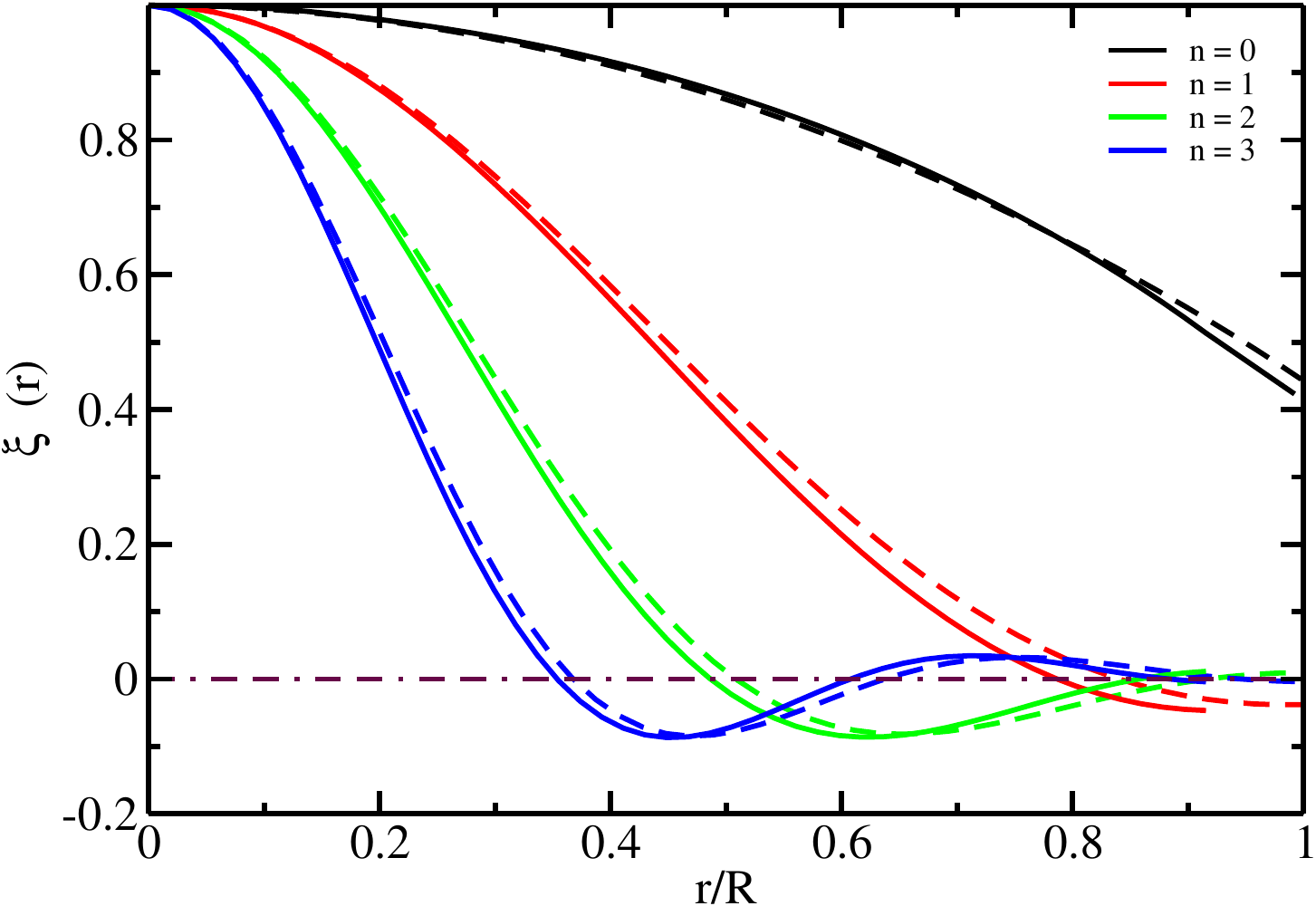}\hfil
.    \includegraphics[width=0.47\linewidth]{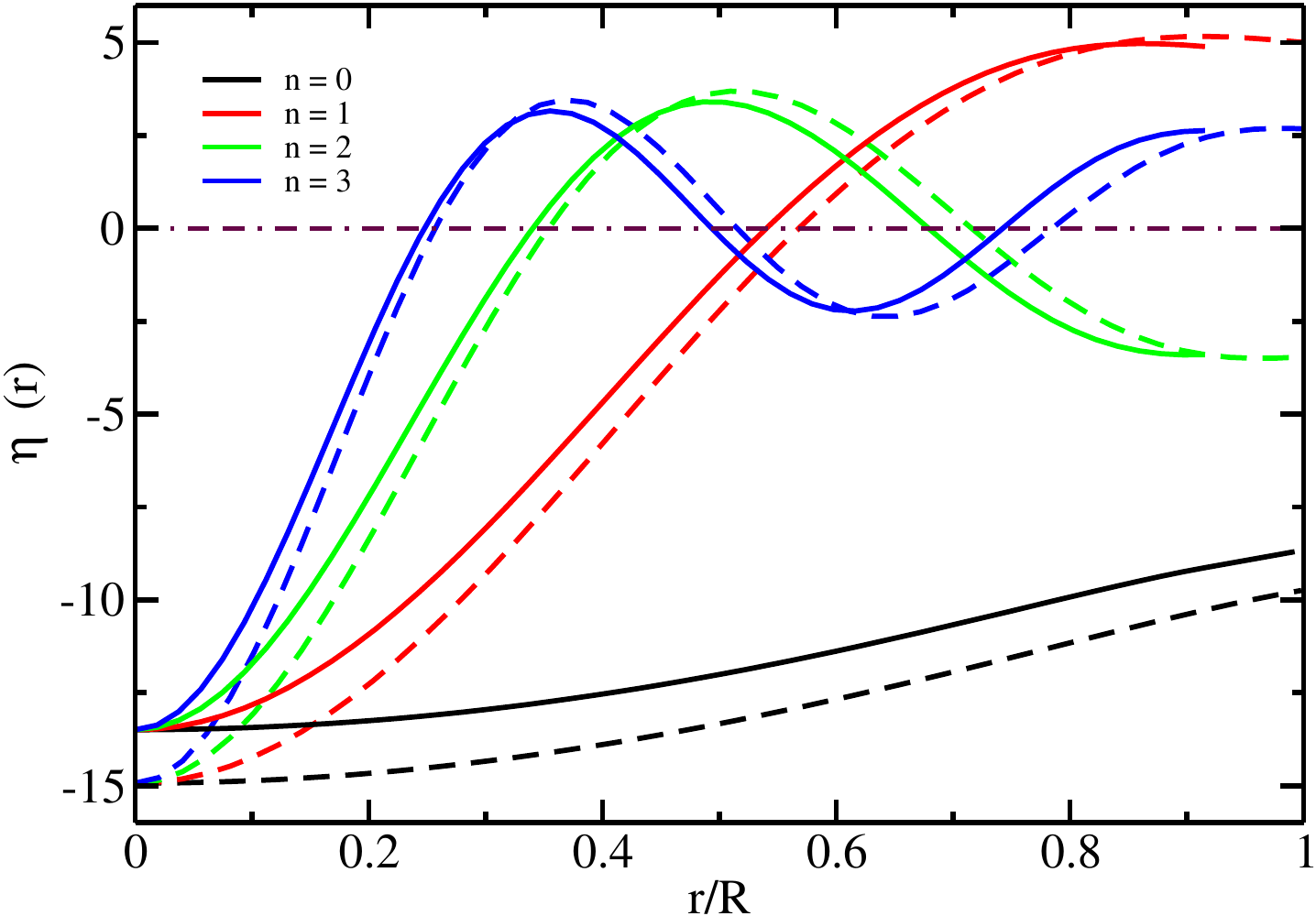}\par\medskip
    \includegraphics[width=0.47\linewidth]{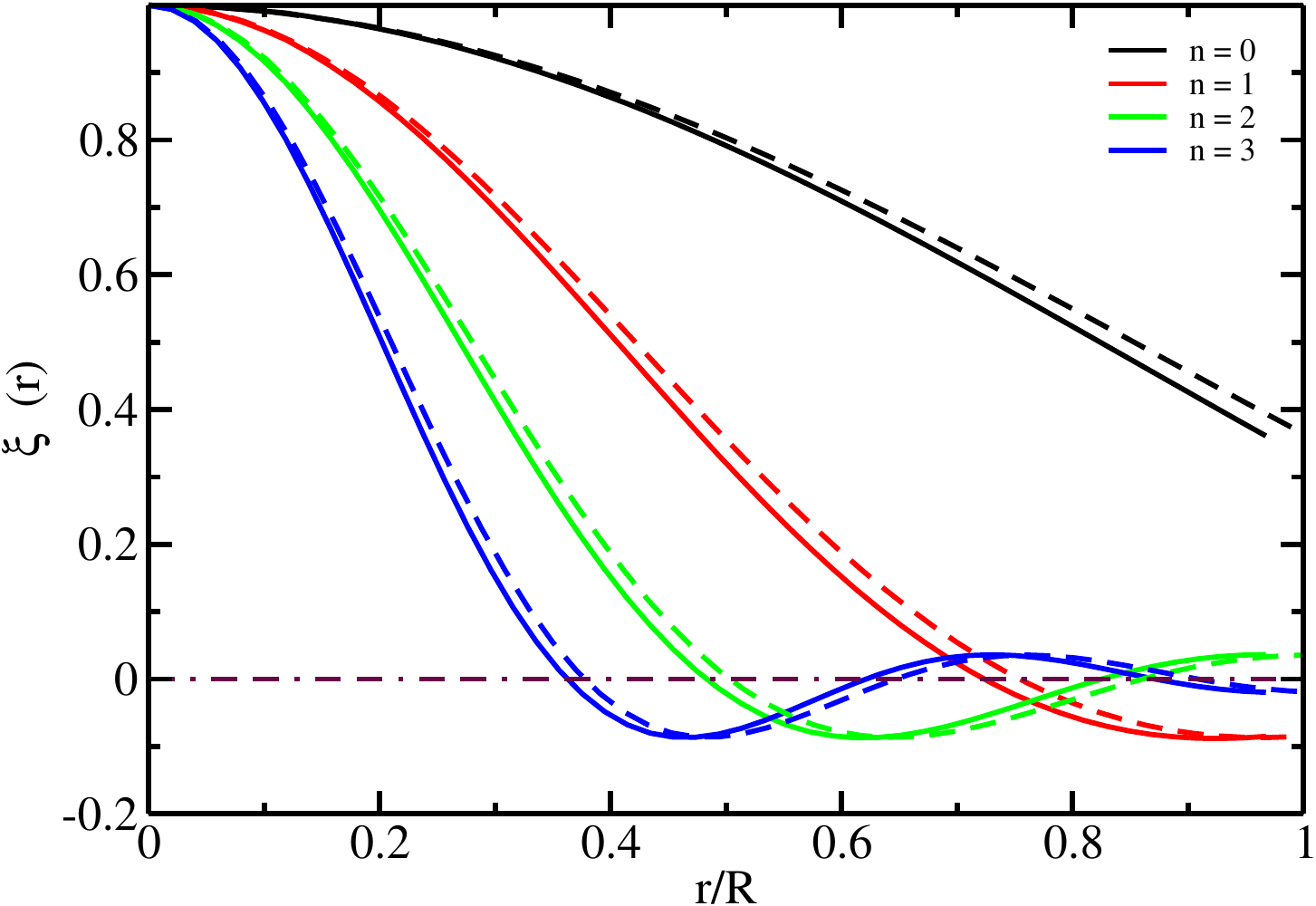}\hfil
    \includegraphics[width=0.47\linewidth]{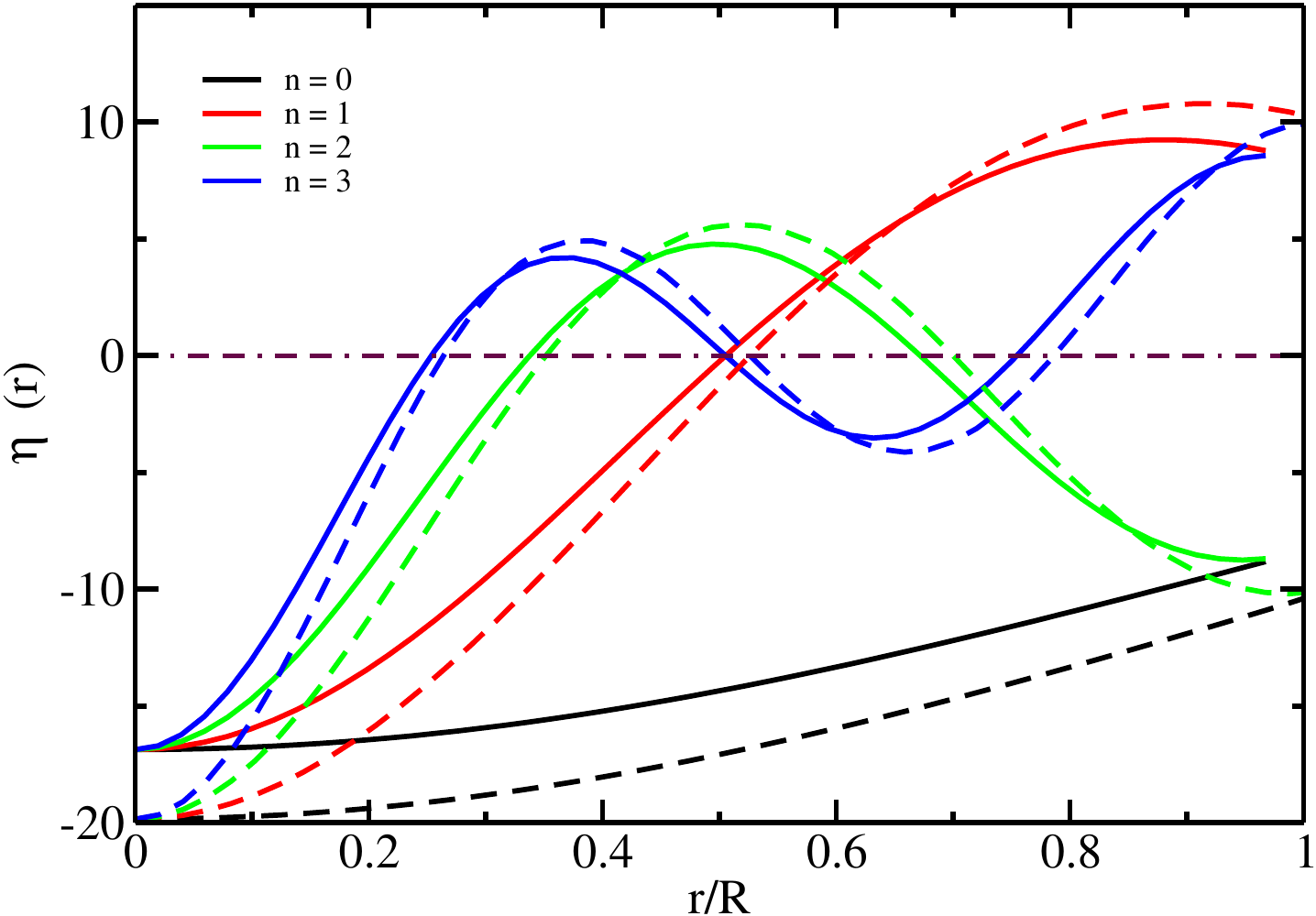}
\caption{ The radial displacement perturbation $\xi(r)$ = $\Delta r/r$ (right panels) and the radial pressure perturbation $\eta(r)$ = $\Delta r/r$ (left panels) as a function of dimensionless radius distance $r/R$ for lower $f$-mode (n = 0), lower order $p$-modes (n = 1, 2, 3). The upper panels represent the result for vBag EoS at $B$ = 55 $\mathrm{ MeV/fm^3}$, $K_{\nu}$ = 9 $\mathrm{GeV^{-2}}$ (solid lines) and $B$= 55 $\mathrm{MeV/fm^3}$, $K_{\nu}$ = 6 $\mathrm{GeV^{-2}}$ (dashed lines), while the lower panels represent the result for the CFL EoS at $B$ = 60 $\mathrm{MeV/fm^3}$, $\Delta$ = 150 $\mathrm{MeV}$, $m_s$ = 150 $\mathrm{MeV}$ (solid lines) and CFL EoS at $B$ = 60 $\mathrm{MeV/fm^3}$, $\Delta$ = 100 $\mathrm{MeV}$, $m_s$ = 0 $\mathrm{MeV}$ (dashed lines).}
\label{fig:radial}
\end{figure*}

The radial displacement perturbation profile $\xi(r)$ and pressure perturbation profile $\eta(r)$ as a function of  dimensionless radius distance $r/R$ for vBag and CFL model EoSs is plotted in Figs.~\ref{fig:radial}. The top panels represent the results  $\xi(r)$ (left panel) and  $\eta(r)$ (right panel) for the vBag model while the lower panels represent the results for the CFL model. Although we have shown only the $f$-mode (n = 0) and lower $p$-modes (n = 1, 2, 3), exactly $n$ nodes are obtained for the $n$th mode both for $\xi$ and $\eta$ profiles, in the region $0<r<R$, thereby following the Sturm-Liouville system. 
The oscillation for $\eta$ is directly proportional to the Lagrangian
pressure variation $\Delta P$, and therefore a decaying amplitude
is observed when it approaches the stellar surface. From Fig.~\ref{fig:radial} (top left panel), one can see that for vBag1 EoS, the amplitude of $\xi_n(r)$ for each frequency mode $\nu_n$ is larger near the center and small at the surface. The lower modes show a smooth drop in their profiles while the higher modes depict small oscillations which would become large for higher modes. For vBag2 EoS, the amplitude of $\xi$ for $n$ = 0  is higher as compared to vBag1 EoS.  The nodes for higher $p$-modes shift towards the center when we use vBag2 EoS meaning when we decrease the coupling constant parameter $K_{\nu}$ from 9 to 6 GeV$^{-2}$.  The system tends to oscillate consistently in a somewhat stable region close to the equilibrium point. For the CFL model, the amplitude of $\xi(r)$ for $n$ = 0 mode is lower as compared to the vBag model EoS. Other modes shift toward the center when we change the superconducting gap parameter from $\Delta$ = 150 to 100 MeV, but the amplitude is higher than the vBag model EoSs.  As one moves toward the star's surface, the amplitude decreases and there is a rapid sign change near the star's center.

\begin{center}
\begin{table}
		\caption{10 lowest order radial oscillation frequencies, $\nu$ in (kHz) for different EoSs considered. For each EoS, the frequencies are calculated at $M$ = $0.77 M_{\odot}$.\label{table1} }
\begin{tabular}{ p{1.0cm}p{1.0cm}p{1.2cm}p{1.2cm}p{1.2cm} }
 \hline
 Nodes & & EoS & & \\
  & vBag1& vBag2 & CFL1 & CFL2\\
 \hline
 0&6.83&7.15&6.47&6.74\\
1	&18.08&18.98&14.18&14.63\\
2	&29.24&30.74&21.61&22.25\\
3	&40.41&42.52&28.96&29.81\\
4	&51.60&54.30&36.30&37.35\\
5	&62.81&66.09&43.62&44.87\\
6	&74.01&77.88&50.93&52.38\\
7	&85.23&89.68&58.24&59.89\\
8	&96.44&101.47&65.56&67.39\\
9	&107.66&113.27&72.87&74.89\\
 \hline
\end{tabular}
\end{table}
\end{center}

\begin{figure*}
\centering
    \includegraphics[width=0.47\linewidth]{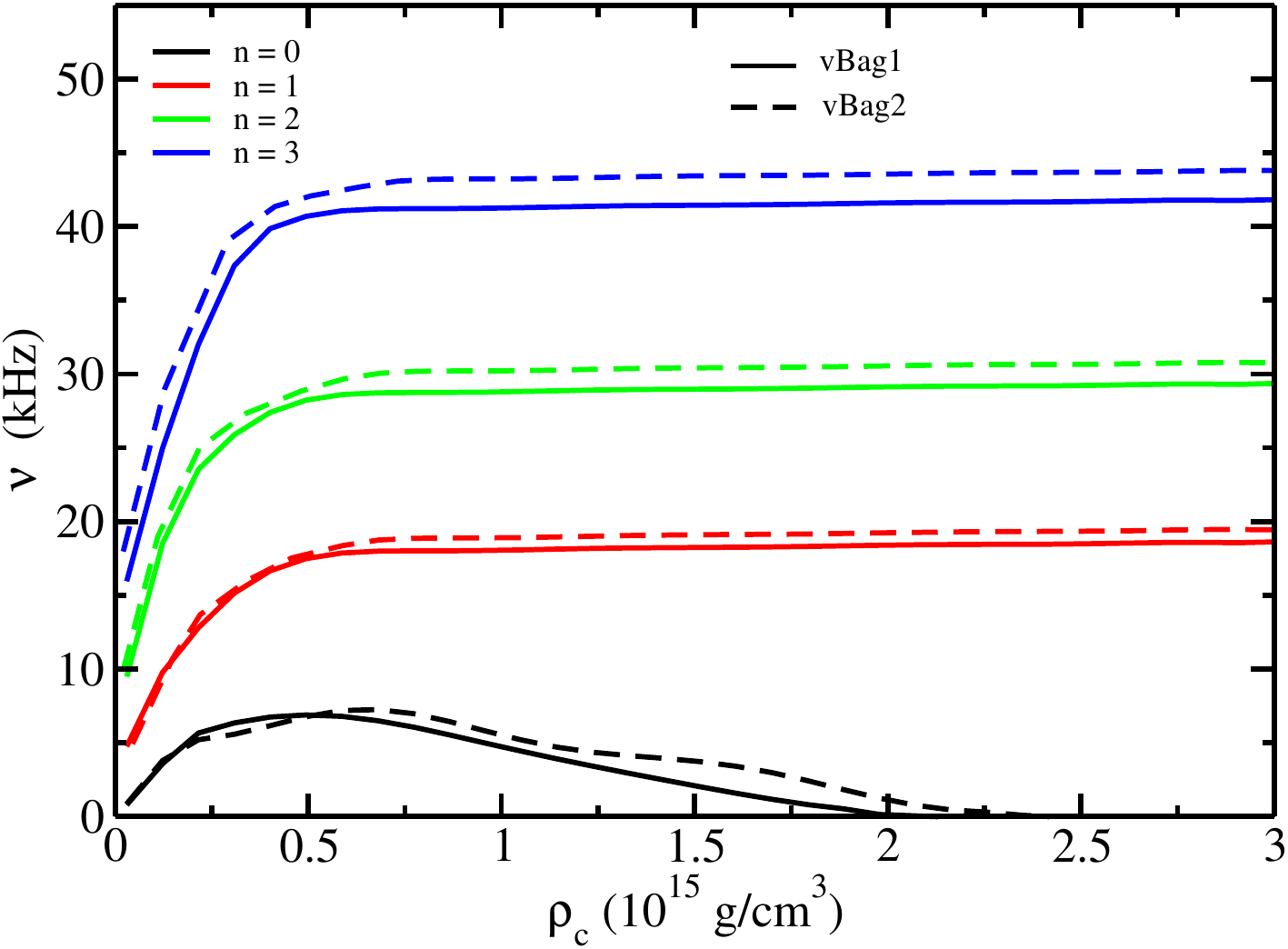}\hfil
    \includegraphics[width=0.47\linewidth]{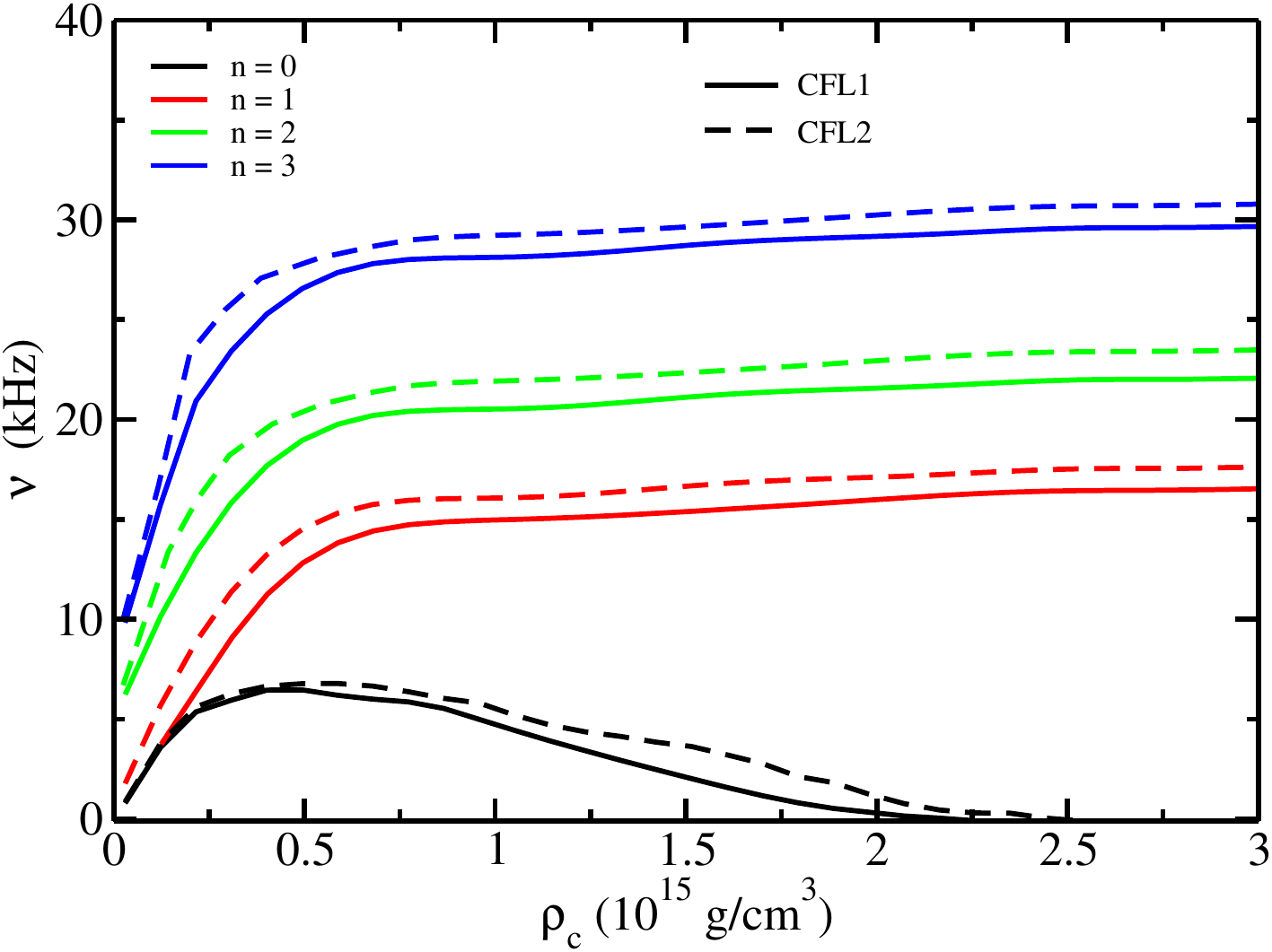}
\caption{ Frequencies of radially oscillating SS as a function of central energy density. The left panel represents the results for vBag EoS at $B$ = 55 $\mathrm{ MeV/fm^3}$, $K_{\nu}$ = 9 $\mathrm{GeV^{-2}}$ (solid lines) and $B$= 55 $\mathrm{MeV/fm^3}$, $K_{\nu}$ = 6 $\mathrm{GeV^{-2}}$ (dashed lines), while the right panel represents the result for the CFL EoS at $B$ = 60 $\mathrm{MeV/fm^3}$, $\Delta$ = 150 $\mathrm{MeV}$, $m_s$ = 150 $\mathrm{MeV}$ (solid lines) and CFL EoS at $B$ = 60 $\mathrm{MeV/fm^3}$, $\Delta$ = 100 $\mathrm{MeV}$, $m_s$ = 0 $\mathrm{MeV}$ (dashed lines). The frequencies for lower radial modes ($n$ = 0, 1, 2, 3) are shown. }
\label{fig:freq_den}
\end{figure*}

From Fig.~\ref{fig:radial}, the pressure perturbation profile for vBag (top right panel) and CFL (bottom right panel) EoSs is shown.
 It is observed that the amplitude of $\eta_n(r)$ is higher both near the star's center and on its surface. For $n$ = 0 mode, the amplitude decreases for vBag2 and CFL2 EoS as compared to vBag1 and CFL1 EoS. The other higher modes for both models shift towards the center at the surface but the amplitude of CFL EoSs is higher than vBag EoSs.
 Despite the fact that the $\eta$ oscillations are directly correlated with the Lagrangian pressure variation $\Delta P$, the $\eta_n(r)$ amplitudes for successive $n$ have large amplitudes near the surface, so the contribution from $\eta_{n+1}$ - $\eta_n$ cancels out because of the opposite signs, satisfying the condition of $P(r = R)$ = 0. This implies that $\eta_{n+1}$ - $\eta_n$  and also $\xi_{n+1}$ - $\xi_n$ are more sensitive to the star's core.  As a result, the measurement of $\Delta \nu_n$ = $\nu_{n+1}$ - $\nu_n$  is an observational imprint of this star's innermost layers. 

 Table \ref{table1} displays the frequencies, $\nu$ in kHz, of the first 10 nodes for all the EoSs considered in this study. All these frequencies are obtained at $M$ = $0.77 M_{\odot}$ for each EoS to determine the $f$- and $p$-mode frequency that the compact object HESS J1731-347 would emit if it's a strange star. The node $n$ = 0 corresponds to the $f$-mode frequency while the others correspond to the lower and highly excited $p$-modes. The frequency for the $f$-mode for vBag2 EoS is higher compared to the rest because of the small maximum mass among all the EoSs.

Fig.~\ref{fig:freq_den} shows the frequencies of radially oscillating SS for vBag (left) and CFL (right) models at different parameter sets, as a function of central energy density for lower radial modes, $n$ = 0, 1, 2, and 3. From the figure, it is evident that stellar models with softer EoSs exhibit higher $f$-mode frequencies than those with stiffer EoSs for the same core density. Larger average densities and more centrally compressed stars are typically associated with the stellar models of softer EoSs. As the center density increases and the $f$-mode frequency ($n$ = 0) simultaneously starts to shift toward zero, the star is getting closer to its stability limit. The stability limit itself exhibits an eigenmode with zero frequency.
 The $f$-mode frequency of vBag model EoSs is higher as compared to the CFL model EoSs because of the lower maximum mass and the corresponding radius that makes the EoS softer. 

For all modes, the frequency appears to decrease or remain constant as the center energy density approaches the minimum value of the specific star model. Higher modes oscillate more frequently than lower stable modes do, and conversely. This is due to the fact that the stars at very high densities can be approximated as being homogeneous and thus the angular frequency $\omega_0^{2}$ follows the relation $\omega_0^{2}$ $\propto$ $\rho (4\gamma - 3)$ \cite{1977ApJS...33..415A,10.1093/mnras/stac2622}. 

\begin{figure}
	\includegraphics[scale=0.32]{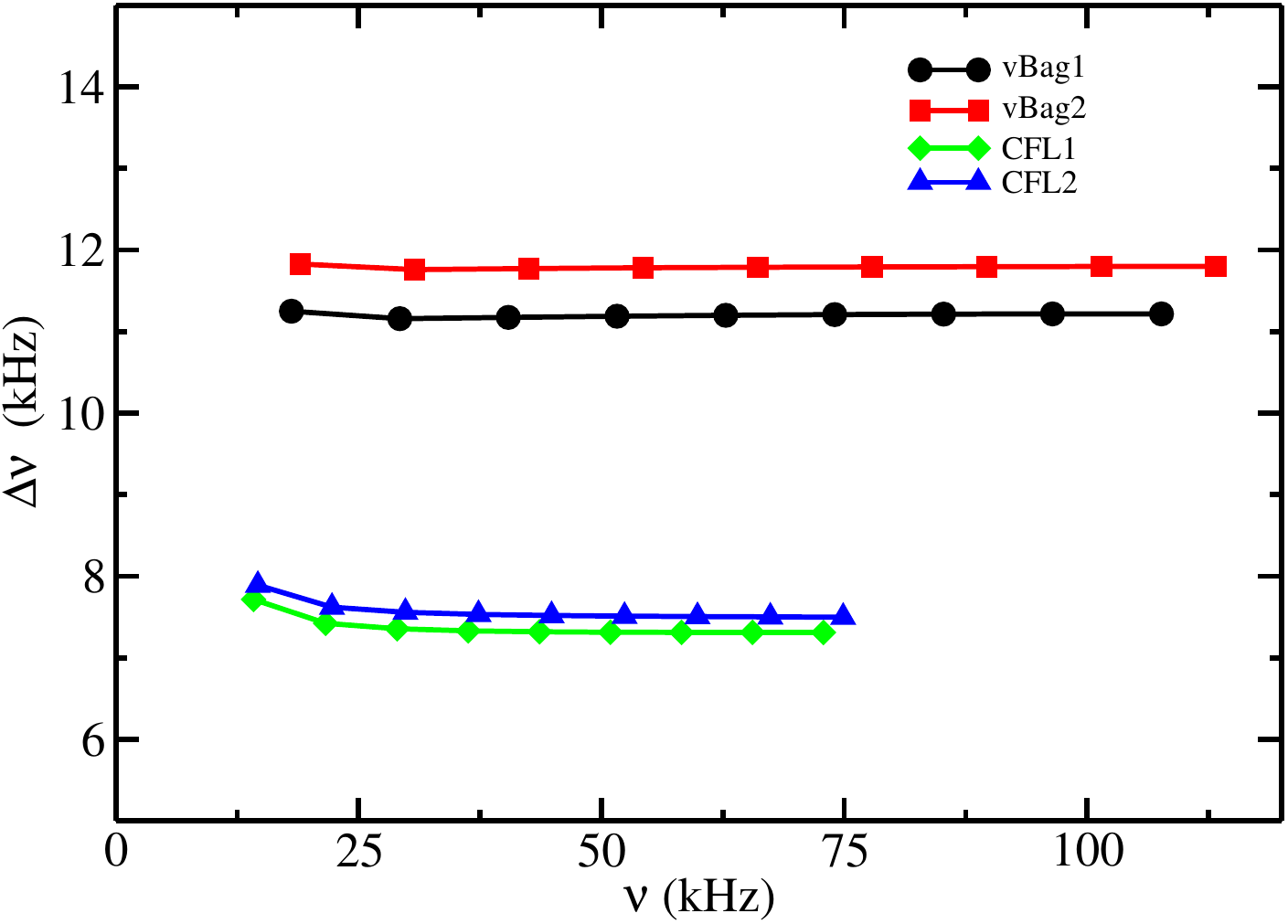}
	\caption{ Frequency difference $\Delta \nu_n$ = $\nu_{n+1}$ - $\nu_n$ vs $\nu_n$ in kHz for vBag and CFL model at different parameter sets.}
	\label{fig:freq} 
\end{figure}

Fig.~\ref{fig:freq} shows the large frequency separation, $\Delta \nu_n$ = $\nu_{n+1}$ - $\nu_n$ vs $\nu_n$, in kHz for vBag1, vBag2, CFL1 and CFL2 EoS. The frequency separation shown here displays all the 10 modes calculated for each EoS as shown in Table~\ref{table1}. Figs. ~\ref{fig:radial} and \ref{fig:freq_den} show only the first 4 modes. For vBag model EoSs, we can see that the frequency separation between consecutive modes shows a smooth trend with a difference of 11 and 12 kHz for vBag1 and vBag2 EoS, respectively. For CFL model EoSs, the frequency difference follows the same trend, although it is smaller for both EoSs than for the vBag EoS, around $(7-8)$ kHz. This is because the vBag model produces softer EoSs, and therefore it oscillates at higher frequencies compared to the CFL model. This shows that the decrease in the central baryon density of the star and, hence, of its mass leads to a large separation $\Delta \nu_n$. We also observe that there are no erratic fluctuations present in $\Delta \nu_n$ in any case. The fluctuations arise from the significant variation of the speed of sound squared $v_s^{2}$ or the relativistic adiabatic index $\gamma$ on the transition layer separating the inner and outer core of the NS, which has an amplitude proportionate to the magnitude of the discontinuity. Also, since we haven't considered the crust EoS, we observe no variations in the speed of sound or relativistic adiabatic index, and so there are no fluctuations.

\section{Summary and Conclusion}
\label{summary}

This study on the peculiar nature of strange stars and their application to the compact object within the supernova remnant HESS J1731-347 holds significant importance in advancing our understanding of this enigmatic astronomical phenomenon. By analyzing different quark models, specifically the vBag model and the CFL model equation-of-states, we aim to comprehend the equation-of-state governing the central compact object (CCO) within HESS J1731-347. 

We employed two different quark models, namely the vBag and CFL models, to study some properties of SSs such as trace anomaly, mass-radius profile, and tidal deformability. Regarding the vBag model, we varied the vector coupling parameter $K_{\nu}$ from (6 - 9) GeV$^{-2}$, and the color superconducting gap parameter $\Delta$ from (100 - 150) MeV, in the CFL model to produce four different EoSs in total. Our aim here was to present the capability of vBag and CFL quark stars to describe the HESS J1731-347 compact object, as well as objects with masses equal to or greater than 2 $M_{\odot}$ limit.

We have studied the trace anomaly $D$ as a function of normalized energy density $\mathcal{E}/\mathcal{E}_0$ for the vBag and CFL model EoSs in different sets of parameters. As far as the vBag model is concerned, the trace anomaly continues to decrease with increasing normalized energy density and drops below 0 at a certain value of normalized energy density for vBag1 and vBag2 EoS. As far as the CFL model is concerned, both CFL1 and CFL2 EoS decrease at lower values of the normalized energy density and then remains almost constant at higher values. We have also studied the speed of sound along with the respective derivative and non-derivative terms, as a function of the normalized logarithmic energy density $D_{log}$ = $ln(\mathcal{E}/\mathcal{E}_0)$. For the vBag model EoSs, the speed of sound shows an increasing behavior and is higher than the conformal limit even at low values of $D_{log}$. $v_{s, nonderiv}^2$ increases with increasing $D_{log}$. The $v_{s, deriv}^2$ shows the opposite behavior to $v_{s, nonderiv}^2$. It continues to decrease up to a certain value of $D_{log}$ and remains almost constant thereafter. For CFL model EoSs, the speed of sound increases with the density but does not violate the conformal limit even at higher densities. 

Using the TOV equations, we observed that all four EoSs satisfy the mass-radius constraints coming from the HESS J1731-347 object, while at the same time, the vBag model EoSs satisfy the mass and radius constraints from coming from other observational data, such as GW170817, PSR J0740+6620, and PSR J0030+0451. The CFL model predicts very stiff EoSs that satisfy the maximum mass of PSR J0952-0607, $M$ = 2.35$^{+0.17}_{-0.17}$ $M_{\odot}$, at $\Delta$ = 100 MeV and the maximum mass of the secondary component GW190814, $M$ = 2.50-2.67 $M_{\odot}$, at $\Delta$ = 150 MeV. The dimensionless tidal deformability  for all four EoSs lies well within the constraint from GW170817, $\Lambda$ = 190$^{+390}_{-120}$.

We have also investigated several modes of radial oscillations of SSs with different EoSs at the mass corresponding to the maximum mass of HESS J1731-347, $M$ = 0.77 $M_{\odot}$, to obtain the frequencies that could be emitted by this CCO, provided that it is an SS. We have studied the 10 lowest eigenfrequencies and corresponding oscillation functions of the vBag model and the CFL model EoSs, solving the Sturm-Liouville boundary value problem and also verifying its validity.

The frequency difference between consecutive modes exhibits a smooth trend without fluctuations. This is due to the fact that we have not included the crust, and so we observe no variations neither in the speed of sound nor in the relativistic adiabatic index. Moreover, in the case of the vBag model, as it produces a softer EoS, the pulsating stars oscillate at higher frequencies compared to the objects modeled by the CFL stiffer EoS.

Regarding vBag1 EoS, the amplitude of radial displacement perturbation $\xi_n(r)$ for each frequency mode $\nu_n$ is larger near the center and small at the surface. The lower modes show a smooth drop in their profiles, while the higher modes depict small oscillations that would become large for higher modes. For vBag2 EoS, the amplitude of $\xi$ for $n=0$ is higher compared to vBag1 EoS. The nodes for higher $p$-modes shift towards the center when we use vBag2 EoS meaning when we decrease the coupling constant parameter $K_v$ from 9 to 6 GeV$^{-2}$. The system tends to oscillate consistently in a somewhat stable region close to the equilibrium point. 

Regarding the CFL model, the amplitude of $\xi(r)$ for the $n=0$ mode is lower compared to the EoS vBag model. Other modes shift toward the center when we change the superconducting gap parameter $\Delta$ from 150 to 100 MeV, but the amplitude is higher than the EoSs of the vBag model. As one moves towards the surface of the star, the amplitude decreases and there is a rapid change of sign near the center of the star. The amplitude of the radial pressure perturbation $\eta_n(r)$ is higher near the star's center and on its surface. For the $n=0$ mode, the amplitude decreases for vBag2 and CFL2 EoS compared to vBag1 and CFL1 EoS. The other higher modes for both models shift towards the center at the surface but the amplitude of CFL EoSs is higher than vBag EoSs.

Contrary to the measurements such as GW170817, NICER, and the heaviest compact stars that examine the characteristics of strongly interacting matter at high densities, the HESS J1731-347 offers crucial information in the density range of 1-2 times the nuclear saturation density, and hence will impose some strict constraints on strongly interacting matter.
By comparing the radial oscillations of the two models and determining the frequencies of the HESS J1731-347 at its maximum mass, the study goes beyond merely outlining the properties of the CCO. This offers a means to investigate their relationship to the observed mass-radius relationship of the HESS J1731-347 compact object and offers insightful information on the dynamical behavior of strange stars. This highlights the fundamental characteristics of strange stars and explains their significance in the context of the current understanding of the HESS J1731-347 compact object by delving into these intricate aspects.

\begin{acknowledgement}
	We wish to thank the anonymous reviewer for useful comments and suggestions
I.~A.~R. and I.~L. acknowledge the Funda\c c\~ao para a Ci\^encia e Tecnologia (FCT), Portugal,
for the financial support to the Center for Astrophysics and Gravitation (CENTRA/IST/ULisboa)
through the grant Project~No.~UIDB/00099/2020  and grant No. PTDC/FIS-AST/28920/2017.    
\end{acknowledgement}


%
%
%

\end{document}